\documentclass[aps,prl,preprint,groupedaddress]{revtex4} 

\usepackage{graphicx}

\newcommand{\mj}{$M_{\mathrm{J}}$}

\newcommand{\me}{$M_{\oplus}$}

\newcommand{\cp}{\citep}
\newcommand{\ct}{\citet}
\newcommand{\apjl}{ApJ}

\newcommand{\aap}{Astron. Astrophys.}

\begin{document}

\title{Frontiers of the physics of dense plasmas and planetary interiors:  experiments, theory, applications}

\author{J. J. Fortney} 
\affiliation{Department of Astronomy and Astrophysics, UCO/Lick Observatory, University of California, Santa Cruz, CA; jfortney@ucolick.org}

\author{S. H. Glenzer}
\affiliation{Lawrence Livermore National Laboratory, L-399, Livermore, CA; glenzer1@llnl.gov}

\author{M. Koenig}
\affiliation{Laboratoire pour l'Utilisation des Lasers Intenses, UMR7605, CNRS-CEA-Universite ParisVI-Ecole Polytechnique, 91128 Palaiseau Cedex, France; michel.koenig@polytechnique.edu}

\author{B. Militzer}
\affiliation{Department of Earth and Planetary Sciences and of Astronomy, University of California, Berkeley, CA; militzer@berkeley.edu}

\author{D. Saumon}
\affiliation{Los Alamos National Laboratory, PO Box 1663, MS F663, Los Alamos, NM; dsaumon@lanl.gov}

\author{D. Valencia}
\affiliation{Department of Earth and Planetary Sciences, Harvard University, Cambridge, MA; valencia@eps.harvard.ed}

\begin{abstract} 
Recent developments of dynamic x-ray characterization experiments of dense matter are reviewed, with particular emphasis on conditions relevant to interiors of terrestrial and gas giant planets.  These studies include characterization of compressed states of matter in light elements by x-ray scattering and imaging of shocked iron by radiography.  Several applications of this work are examined.  These include the structure of massive ``Super Earth'' terrestrial planets around other stars, the 40 known extrasolar gas giants with measured masses and radii, and Jupiter itself, which serves as the benchmark for giant planets.
\end{abstract}

\maketitle

\section{I. Introduction}
We are now in an era of dramatic improvement in our knowledge of the physics of materials at high density.   For light elements, this theoretical and experimental work has many applications, including internal confinement fusion as well as the interiors of gas giant planets.  For heavy elements, experiments on silicates and iron at high pressure are helping to better understand the Earth, as well as terrestrial planets as a class of objects.  In particular, the discovery of rocky and gaseous planets in other planetary systems has opened our imaginations to planets not found in our own solar system \cp{Charb06}.

While the fields of experiments of matter at high densities, first principles calculations of equations of state (EOS), planetary science, and astronomy do progress independently of each other, it is important for there to be communication between fields.  For instance, in the realm of planets, physicists can learn of key problems that exist in the area of planetary structure, and how advances in our understanding of input physics could shed new light in this area.  Astronomers and planetary scientists can learn where breakthroughs in physics of materials under extreme conditions are occurring, and be ready to apply these findings within their fields.

This brief review focuses on work presented at the joint American Physical Society (APS), High Energy Density Laboratory Astrophysics (HEDLA), and High Energy Density Physics (HEDP) meeting in April, 2008.  We first discuss some experimental and theoretical work on light elements, including some applications to gas giant planets, which are predominantly composed of hydrogen and helium.  We discuss new models of the interior structure of Jupiter and review the observed mass-radius relationship of extrasolar giant planets (EGPs).  We then focus on terrestrial planets and investigate new experimental work on iron.  We then discuss the predicted structure of ``Super Earth'' planets, predominantly solid planets made up of iron, rock, and water, from 1-10 Earth masses (\me).

\section{II. Light Elements and Giant Planets}
\subsection{A. Pulsed x-ray probing of light elements}
For accurate measurements of densities and temperatures in dense and compressed matter, novel x-ray Thomson scattering techniques\cp{Landen01a} have been developed.  These experiments employ powerful laser-produced x-ray sources that penetrate through dense and compressed materials with densities of solid and above.  Both thermal Ly-$\alpha$ and He-$\alpha$ radiation from nanosecond laser plasmas\cp{Landen01b} or ultra short pulse laser-produced K-$\alpha$ inner-shell emission\cp{Kritcher07} have been shown to fulfill the stringent requirements on photon numbers and bandwidth for spectrally resolved x-ray scattering measurements in single shot experiments.  Experiments have been performed in the non-collective (backscatter) regime, where the scattering spectrum yields the Compton feature\cp{Glenzer03}. On the other hand, collective scattering on plasmons, i.e., electron density (Langmuir) oscillations, have been observed in forward scatter geometry\cp{Glenzer07}.

These techniques have recently been applied to shock-compressed beryllium (H.~J.\ Lee et al., in prep).  In experiments at the Omega laser facility\cp{Sawada07} twelve laser beams (500 J each) directly illuminate the foil with laser intensities of $I = 10^{14} - 10^{15}$ W cm$^{-2}$ producing pressures in the range of 20-60 Mbar and compressing the foil by a factor of 3.  The Compton scattering spectrum of the 6.18 keV Mn He-$\alpha$ and 6.15 keV intercombination x-ray probe lines measured at $\Theta$ = 90$^o$ scattering angle shows a parabolic spectrum downshifted in energy from the incident radiation by the Compton effect; the shift is determined by the Compton energy $E_C = \mathrm{h^2 k^2}/4\pi m_e = 74$ eV, with $k = 4 \pi (E_0/ \mathrm{hc}) \sin(\Theta / 2) = 4.4 \AA^{-1}$, and $E_{\rm 0}$ the energy of the incident x-rays.   The Compton scattering spectrum directly reflects the electron distribution function; for a Fermi-degenerate system the width of the Compton spectrum provides the Fermi energy, $E_F \sim n_e^{2/3}$. Unlike for plasmas with a Maxwell-Boltzmann distribution, the width is sensitive to the electron density.  In addition, the intensity ratio of the elastic to inelastic scattering feature from Fermi-degenerate plasmas is sensitive to the ion temperature because elastic scattering is dependent on the ion-ion structure factor.  

\begin{figure}
\includegraphics[scale=1.0]{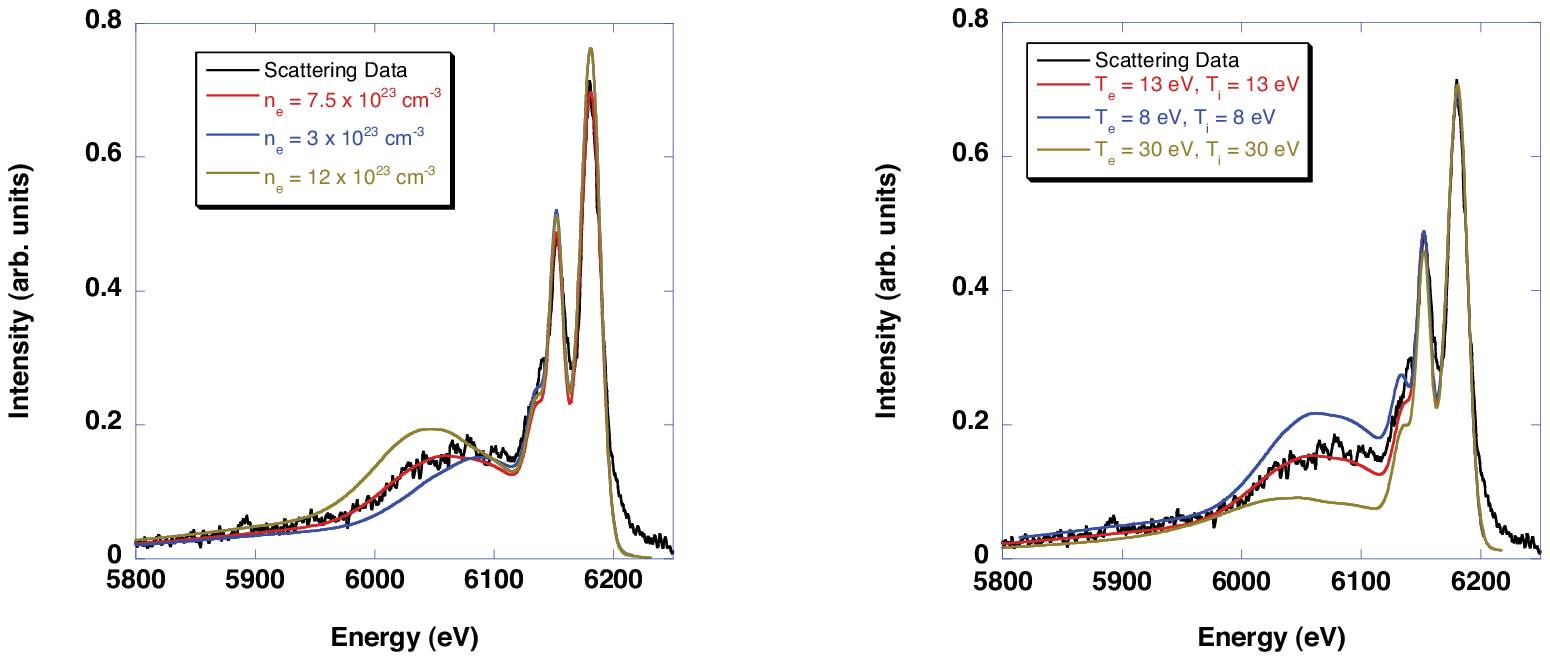}
\caption{(Color online) Compton scattering spectrum from laser compressed Fermi-degenerate beryllium. The width of the Compton feature is sensitive to density and its relative intensity is sensitive to the ion temperature.
\label{scat}}
\end{figure}

Figure (\ref{scat}) shows the scattering data along with calculated scattering spectra for which the electron density (left) and the temperature (right) has been varied.  For the analysis we assume $T_e$=$T_i$ and $Z$ = 2 consistent with calculations and with the measurements from isochorically heated Be.  Density and temperature obtained in this way are $n_e = 7.5 \times 10^{23}$ cm$^{-3}$ and $T$ = 13 eV for representing a Fermi temperature of $E_F$ = 30 eV and scattering parameter $\alpha$ = $1/(k {\lambda}_S) = 0.48$.  The error bar for the measurement is of order $<10$\%, dominated by noise.  This error estimate is not affected by uncertainties of $Z$ because the shape of the Compton scattering profile provides an additional constraints.  The parameters inferred from the theoretical fit match radiation hydrodynamic simulations of this experiment to 10\% and also agree with the results of forward scattering measurements that have independently measured $n_e$ and $T_e$ from the plasmon spectrum.

The experiments have directly measured the conditions and dynamic structure factors of shocked matter, thus going beyond characterization of shock wave experiments with particle and shock velocities.  These novel experiments have only now become possible with the advent of penetrating powerful x-ray probes produced on high-energy density physics facilities. This feature further allows testing of radiation-hydrodynamic calculations with different EOS models for shock-compressed matter.  Future experiments will apply Compton scattering to measure the compressibility and adiabat of compressed matter, including hydrogen \cp{Holl07}.

\subsection{B. Experiment and Theory of H/He}
\begin{figure}[t]
\includegraphics[width=\columnwidth, scale=0.8]{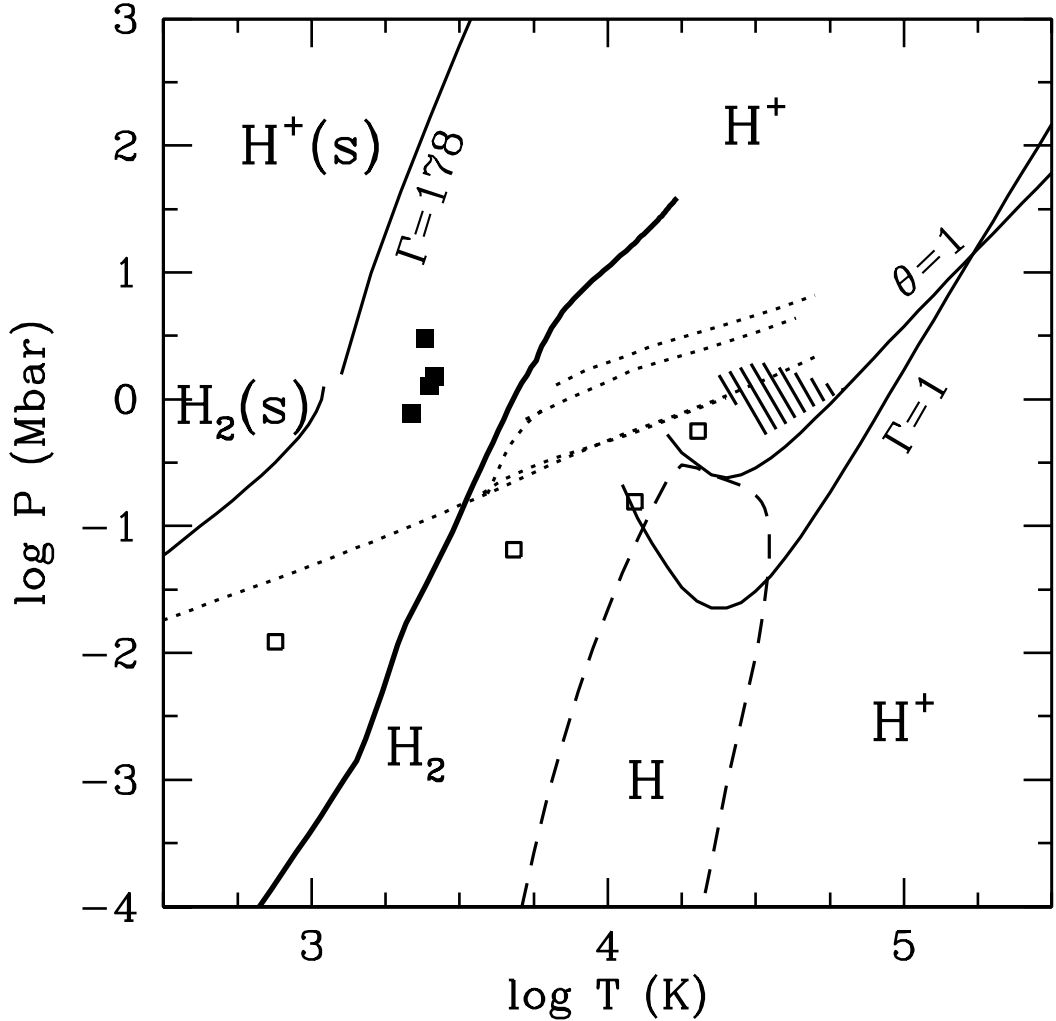}
\caption{Phase diagram of hydrogen. Physical regimes are indicated with solid lines showing the plasma coupling parameter ($\Gamma=1$) and the electron degeneracy parameter ($\theta=1$). For $\Gamma > 178$, the plasma freezes into a bcc Coulomb solid.  The melting curve of H$_2$ is also indicated for $\log T \le 3$.  The dashed curve shows the dissociation and ionization boundaries in the low density gas.  Above $\log P \sim 0.7$, hydrogen is fully ionized.  The Jupiter isentrope is shown by the heavy solid line.  The regions probed by single, double and triple shock experiments on deuterium are indicated with dotted lines.  Filled squares show the near isentropic compression data of \citet{fortov07} suggesting a PPT in hydrogen.  The single and double shock helium points of \citet{Nellis84} are indicated with open squares.  Finally, the hashed region outlines the locus of the shocked states achieved by \citet{eggert08} from pre-compressed He targets.}
\label{fig:phaseD}
\end{figure}

Experiments on deuterium and helium provide vital EOS data to model
the interior of giant planets (Fig. \ref{fig:phaseD}). Recent work include
explosively driven shocks in hydrogen and deuterium \citep{fortov07} that provided
experimental evidence of pressure-driven dissociation transition in
dense hydrogen that is associated with a rapid increase in
conductivity \cp{Weir96,fortov07}. The results in \ct{fortov07} were interpreted as signs of
a plasma phase transition of first-order. However, within experimental
error bars, an interpretation in terms of a gradual dissociation
transition as predicted from first-principles simulations \cp{Militzer08} is also
possible.

\begin{figure}
\includegraphics[width=\columnwidth, scale=0.85]{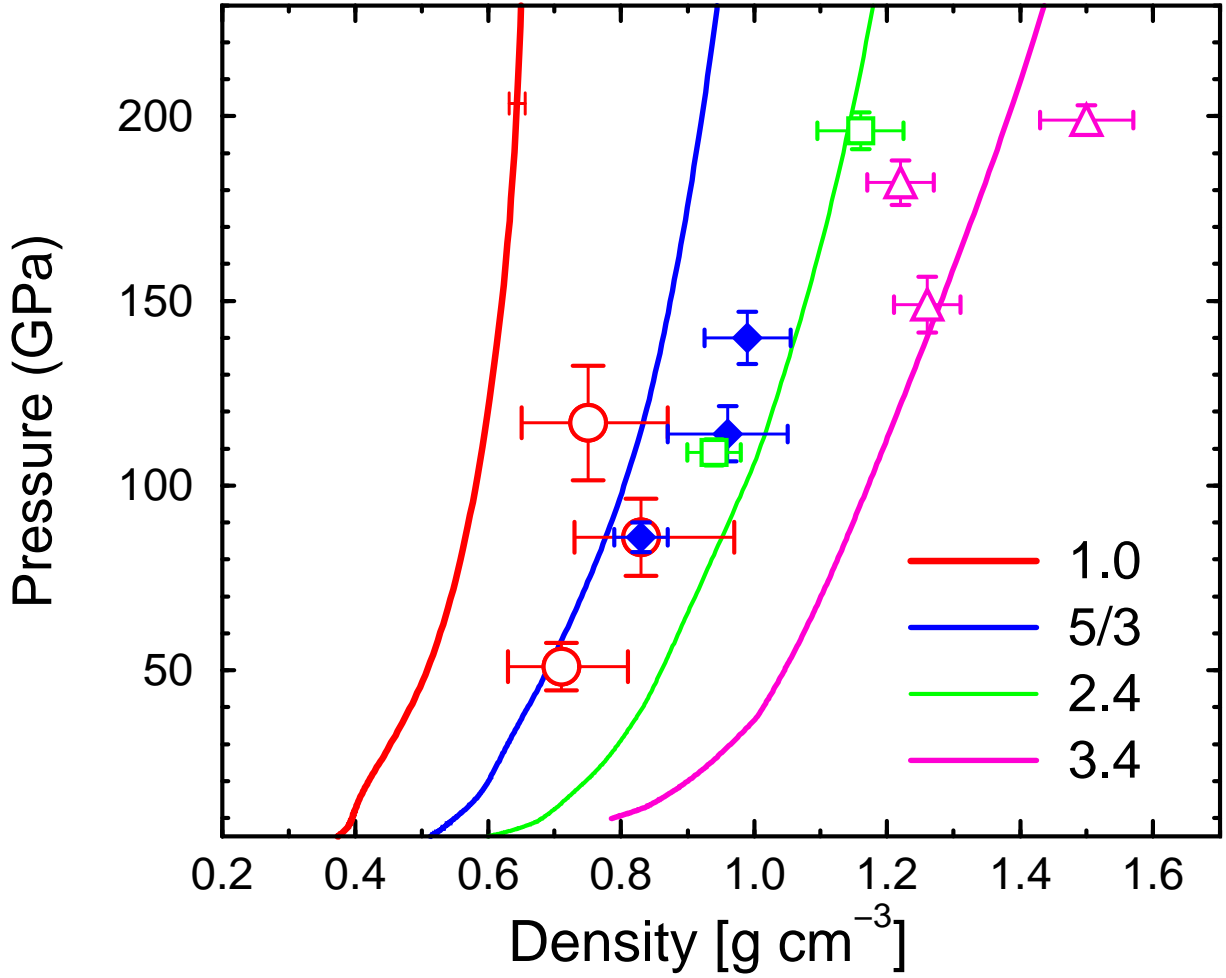}
\caption{(Color online) Comparison of between theory (solid lines) and laser shock wave experiments (symbols). Helium was exposed to extreme temperatures and pressure that are relevant for planetary interiors. The colors represent different precompression ratios. The ability to precompress samples statically before launching the shock is an important experimental improvement that allows to probe deeper in the giant planet interiors. Good agreement between theory and experiment is found for the higher precompressions.
\label{He}}
\end{figure}

While standard shock wave experiments can reach very high temperatures
and megabar pressures, they increase the sample density only 4 to 5
fold~\cite{MH07}. There is consequently very little data taken
directly under the conditions found inside Jupiter and Saturn. In
addition to existing isentropic and off-hugoniot shock compression
techniques, this issue is addressed with a new experimental method
that combines static and dynamic compression techniques. By
precompressing the sample statically in a modified diamond anvil cell
\cp{eggert08}, a higher final shock density is reached, and one can therefore
probe deeper in the planetary interiors. In Figure (\ref{He}) we show a
comparison of the \citet{eggert08} precompressed data (the hashed
region in Fig. \ref{fig:phaseD}) to first principles simulations of \citet{Militzer08b}. The agreement is particularly good at high
compression but deviations are observed for the measurements without
precompressions. Further experimental and theoretical work will be
needed to resolve this discrepancy.

\subsection{C. Jupiter and Extrasolar Giant Planets}
An important requirement of theories of planet formation is to account for the present day structure of Jupiter and Saturn, which are relatively well observed. The history of their formation is imprinted primarily in the amount and distribution of heavy elements in their interior. Heavy elements are supplied as solid bodies while the massive H/He envelopes of giant planets accumulate through gas accretion.  The relative importance of the two accretion processes during the formation of the planet is closely tied to the formation process and the surrounding environment. As noted by \ct{Saumon04}, uncertainties in H/He EOS dominate all uncertainties when trying to understand the interior structure of Jupiter.  It is then essential that accurate EOS measurements can be made for light elements under giant planet conditions.

\begin{figure}
\includegraphics[scale=1.0]{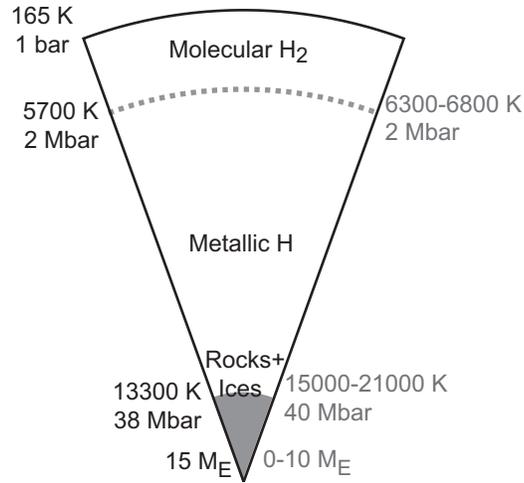}
\caption{Schematic interior view of Jupiter, based on \ct{Militzer08}.  Running along the left in black are pressures and temperatures from their model at three locations, as well as the core mass estimate ($\sim$15\me).  The transition from molecular hydrogen (H$_2$) to liquid metallic hydrogen (H$^+$ is continuous.  Running along the right in gray are these same estimates from \ct{Guillot05}.
\label{js}}
\end{figure}

Interestingly, the first two modeling efforts based on first principles EOS computed independently but with essentially the same method give very different results
for the amount and distribution of heavy elements in Jupiter \citep{nettelmann08,Militzer08}.  The reasons for this discrepancy are discussed in \citet{mh08} and stem primarily from different assumptions  for the interior structure of Jupiter.  \citet{nettelmann08} assumed a different concentration of heavy elements and helium for the molecular and the metallic regime that could for example be introduced by a first order phase transition in hydrogen. \ct{Militzer08} found no evidence of such sharp transition in their first-principle simulation and concluded the mantle must be isentropic, fully convective, and of constant composition.  Jupiter's interior structure, as derived by \ct{Militzer08}, is shown in Figure (\ref{js}).  We may be at the threshold where the EOS of H/He mixtures is understood well-enough to force qualitative changes in our picture of the basic structure of Jupiter.  The implications for the interior of Saturn and, by extension, for the planet formation process remain to be explored.

New observational data on Jupiter is hard to come by, given that space probes are necessary to measure deep atmospheric abundances and map the gravitational and magnetic fields.  In 2011 NASA will launch the JUNO orbiter, which will reach Jupiter in 2016 \cp{Bolton06}.  This orbiter has several important goals relating to the structure of the planet.  The deep abundance of oxygen will be measured, which is potentially Jupiter's third most abundant element, after H and He.  The detailed mapping of Jupiter's gravity field will give us unparalleled access into the internal structure of the planet \cp{Hubbard99b}.

\begin{figure}
\includegraphics[width=\columnwidth, scale=1.0]{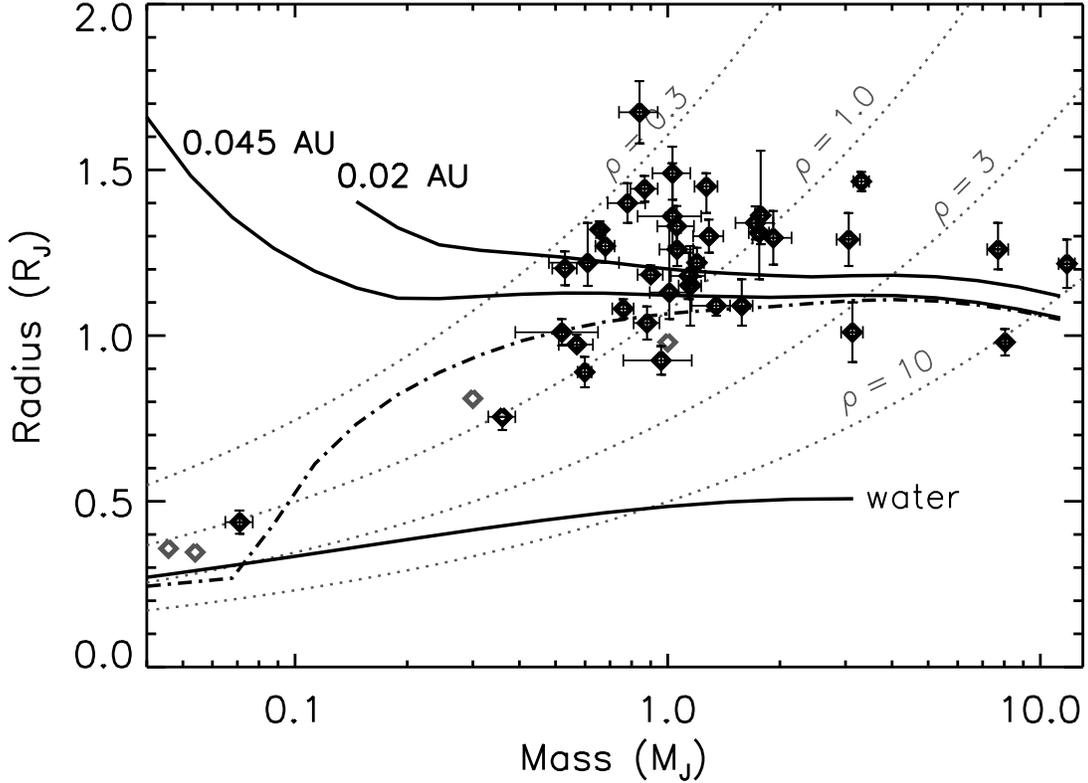}
\caption{A comparison of theoretical mass-radius curves for gas giant planets and 40 observed transiting planets, using the models from \ct{Fortney07a}.  The majority of these planets orbit at distances of only 0.02 to 0.05 AU from their parent stars, while the Earth orbits at 1 AU (by definition).  The x-axis is mass in Jupiter masses, and the y-axis radius in Jupiter radii.  The top two solid black curves are for pure H-He, 4.5 Gyr-old, giant planets at 0.02 AU and 0.045 AU from the Sun.  (The Earth-Sun distance is 1 AU.)  The thick dash-dot curve also shows models at 0.045 AU, but with 25 \me\ (0.08 \mj) of heavy elements (ice+rock) in a core.  A mass-radius curve for pure water planets is also shown.  Gray diamonds are, left to right, Uranus, Neptune, Saturn, and Jupiter.  Black diamonds with error bars are the transiting planets.  Curves of constant bulk density (in g cm$^{-3}$) are overplotted in dotted gray.
\label{mr}}
\end{figure}

The available data set on transiting EGPs continues to expand.  These planets periodically pass in front of their parent stars, allowing for a determination of planetary radii.  Planetary masses are determined from the Doppler shift of the parent star's spectral lines.  Forty such planets, most in very close-in orbits, are now known around other stars.  The doubling time for the number of detected planets is now \emph{less than one year} \cp{Charb08b}.  Dedicated space missions such as CoRoT \cp{Borde03,Barge08} and Kepler \cp{Basri05} will detect dozens to hundreds of additional Neptune-like to Jupiter-like planets, in additional to smaller terrestrial planets, discussed below.  In Figure (\ref{mr}) we show the measured masses and radii of known planets, compared to the solar system's giant planets, and predictions from theoretical models at two different irradiation levels\cp{Fortney07a}.

It is the extreme diversity shown in Figure (\ref{mr}) that is most surprising.  Irradiated giant planets were expected to be inflated relative to Jupiter\cp{Guillot96}, but the range of radii for planets of similar masses does not yet have a satisfactory solution.  To explain the relatively smaller radii of some of the Jovian planets, heavy element abundances of 100-200 \me\ (0.31-0.62 \mj) are needed.  However, we are ignorant of whether these heavy elements are predominantly mixed into the H/He envelope or within a distinct core.  This issue is actually even more complicated, due to several additional factors\cp{Baraffe08}:  (i) the differences between the various equations of state used to characterize the heavy material (water, rock, iron), (ii) the chemical composition of the heavy elements (predominantly water or rock?), and (iii) their thermal contribution to the planet evolution (which is often ignored altogether).  Deriving heavy element abundances based on a given planet's mass and radius will be uncertain, given these issues.  As we discuss in \S III, similar issues are also important for massive terrestrial planets.  

To explain the large radii of many of the planets, either an additional internal energy source must be invoked,\cp{Bodenheimer01,Guillot02} or that the cooling and contraction of these planets has been stalled.\cp{Burrows07,Chabrier07c}  The only clear trend to date is that planets around metal-rich parent stars tend to possess larger amounts of heavy elements.\cp{Guillot06}

\section{III. Heavy Elements and Terrestrial Planets}
\subsection{A. Experiments on Iron}
With current technology, diamond cell experiments (static), do not allow one to obtain meaningful data at temperatures of several thousands of K once the pressure exceeds 200 GPa. Although measurements can be achieved with good precision below 200 Gpa \cp{Boehler93}, the melting curve of iron or iron alloys at the inner core boundary (330 GPa, about 5000 K) is thus beyond the capabilities of these experiments. On the other hand, dynamic experiments can easily reach inner core pressures but the corresponding temperatures, which are large and are fixed by the Hugoniot curve, do not allow one to explore the relevant \emph{P-T} space. As a result, the iron phase diagram at conditions corresponding to the Earth's inner core has never been directly measured and large uncertainties remain regarding its equation of state (EOS). These unknowns severely limit current Earth modelling as the iron EOS is of utmost importance to constrain the chemical composition and energy balance of the Earth's core. The discovery of low-mass planets outside the solar system renders the exploration of iron at $> 500$ GPa (5 Mbar) pressures and $\sim$1 eV temperatures even more pertinent.

The French National Research Agency (ANR) recently funded a several-year program focused on the development of new diagnostics to study the physical properties of iron, the development of methods to explore broader regions of the EOS diagram, and the combined use of experimental and theoretical methods to characterize the high pressure phases of this element.

The first part of this project is to develop adequate x-ray sources both to radiograph and/or perform diffraction measurements on shock compressed iron. Results at lower energies have already been obtained for aluminum \cp{Ravasio08}.
To this aim, experiments were performed on the 100 TW laser system at LULI, France, which delivers 20 J in 0.3-10 ps at a wavelength of 1057 nm and 6J when frequency-doubled. The latter was used to look at the effects of preformed plasma due to the laser `pedestal' which is 500 ps wide with a contrast of $10^{-6}$.
Different target materials and geometries, as well as the effects of laser parameters and filtering/shielding of the detector were studied. At very high laser intensities, x-rays are generated by energetic electrons produced by the laser-plasma interaction which penetrates the target and produces K-$\alpha$ radiation. The x-ray emission stops a few ps after the end of the laser pulse, as the electrons lose their energy due to classical charged-particle stopping processes. For laser pulses of $\sim$10 ps, a temporal resolution of less than 20 ps can then be achieved. This duration is short enough to resolve shockwaves for density measurements in EOS studies, as the shock velocities are of the order of a few tens $\mu$m ns$^{-1}$.

\begin{figure}
\includegraphics[scale=1.5]{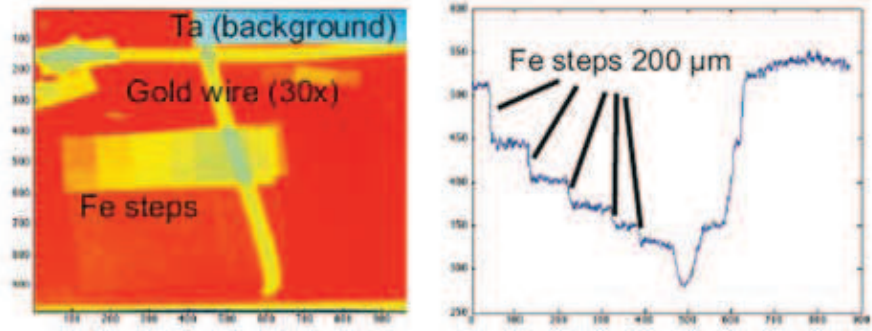}
\caption{(Color online) Resolution tests obtained with 60 keV x-rays produced by a short pulse irradiated W-wire. 
\label{xray1}}
\end{figure}

The backlighter target was made of W, (producing K$\alpha$ energy of 60 keV). To produce a small source size, required for high 2D spatial resolution, we used thin (18 $\mu$m diameter) W wires \cp{Park06}. The spatial resolution of the x-ray source was measured using a crossed pair of 100 $\mu$m diameter gold wires. The latter was used for these high energy x-rays, as the absorption of a standard gold grid was too low ($\sim$5\%). The magnification of this point-projection system was 30$\times$, with a source-detector (imaging plate filtered with Tm) distance of 30 cm; an additional 2.5 cm plastic layer was used to stop energetic charged particles emerging from the target. To characterize the spectral distribution of the x-rays, a transmission crystal spectrometer (DCS) \cp{Seely06}, was implemented in order to measure the contribution of high-energy x-ray background to the radiograph image. The experiment reliably delivered high quality radiographs of static targets with best results obtained at intensities of $10^{18}$ W cm$^{-2}$. Figure (\ref{xray1}) shows a radiograph at 60 keV of a test target (100 $\mu$m thick gold wire at 30$\times$ magnification) obtained from a 18 $\mu$m diameter W-wire target. Analyzing the absorption profile of the wire on the detectors shows a spatial resolution of better than 20 $\mu$m. The contrast on iron steps (right part of Figure \ref{xray1}) shows the resolved density gradients that allows for the deduction of the density of shocked iron with error bars lower than 10 \%.

\begin{figure}
\includegraphics[width=\columnwidth]{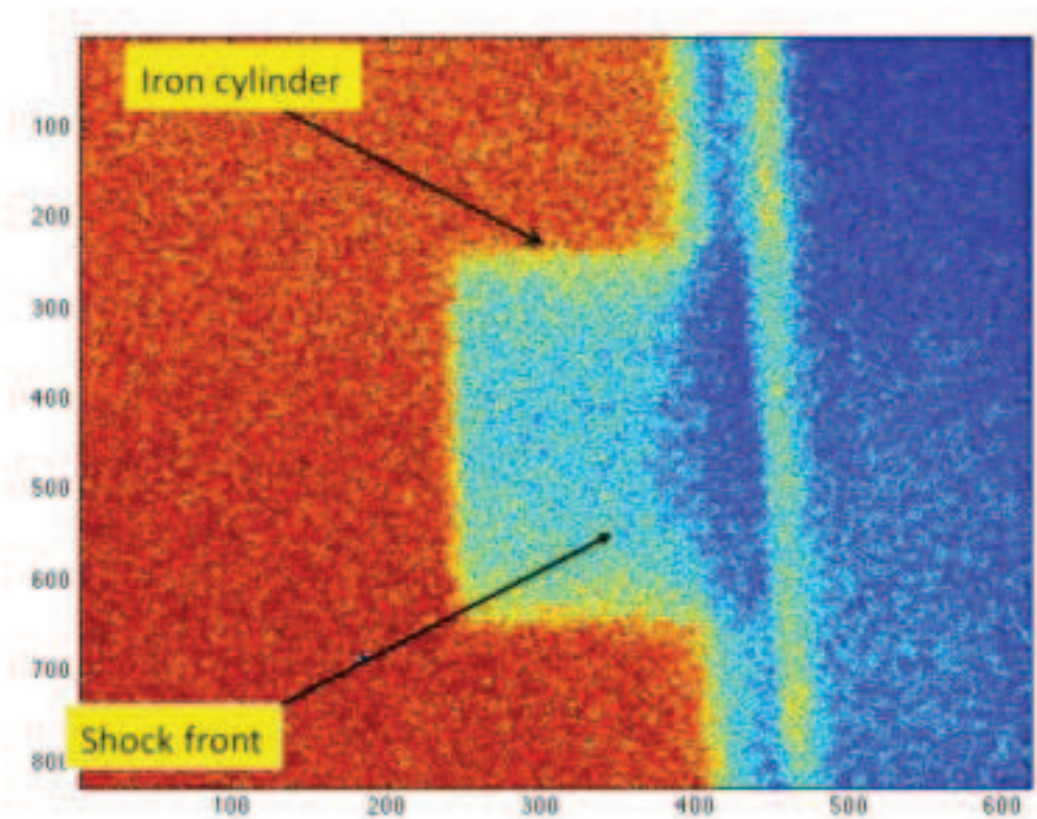}
\caption{(Color online) 60 keV radiograph of a shock compressed iron target.
\label{xray2}}
\end{figure}

This technique has then been applied to radiograph a shock-compressed target made of a ablator pusher and a 500 $\mu$m diameter iron disk, 250 $\mu$m thick. For this experiment, besides the short pulse beam, a high energy long pulse beam was needed to drive a uniform planar shock. Therefore we used the new LULI2000 facility which has this capability and  obtained the first radiograph of a laser shock compressed iron target (Figure \ref{xray2}). Due to a lower contrast on the LULI2000 facility than the 100 TW, the signal/noise ratio is not optimum but the shock front is clearly observable.  Detailed analysis is still underway.

\subsection{B. Application: Super Earths---Massive Terrestrial Planets}
Super-Earths are the newest class of discovered extra-solar planets. These 1-10 earth-mass (\me) planets are likely to consist of solids and liquids rather than of gases. With their relatively large masses, they experience very large internal pressures. Pressure constrains the power law relationship between mass ($M$) and radius ($R$) of solid planets. The value for the exponent in $R = R_{\rm REF} (M/M_{\rm REF })^{\beta}$ is $0.262 \leq \beta \leq 0.274$ as constrained by the different internal structure models for super-Earths, while it is $\beta = 0.3$ for planets between 5-50\% the mass of Earth \cp{Valencia06,Valencia07}. $R_{\rm REF}$ is the radius of a planet with reference mass $M_{\rm REF}$, usually Earth's, that may be rocky or have large amounts of H$_2$O with a correspondingly larger $R_{\rm REF}$. The central pressure of rocky super-Earths (up to $\sim$60 Mbar) scales proportionately with mass, reaching values that challenge the understanding of rock behavior under such extreme conditions. Despite the different treatments in the models \cp{Fortney07a,Seager07,Sotin07} and intrinsic uncertainties in the equation of state (EOS), composition and temperature structure, the mass-radius relationship is robust, and thus, useful for inferring the expected signal in searches for transiting super Earths.

However, information on the structure, such as the size and state of the core, crucially depends on the exact behavior of super-Earth materials (silicates, iron, iron alloy and ices) at high pressures and temperatures. In order to accurately describe the physical properties of super-Earths, such as their ability to have a magnetic field by having a molten core, or to extract information to constrain formation models such as from the existence of a metallic core, we need a very detailed description of super-Earths' interior, that can not be done without improvements in the EOS of silicates, iron alloys and ices.

A few questions that, if addressed, will considerably improve the internal structure models and thus, our interpretation of the data are: (1) What is the stability field of post-perovksite and are there other higher pressure silicate phases? Our lack of knowledge of other existing phases means that the radius in models is an upper value. (2) At the pressure range of super-Earths (up to 60 Mbars), which existing EOS is more accurate? A few high pressure experiments can illuminate the extrapolation qualities and deficiencies of the different EOS used by the models (Vinet, Birch-Murnaghan, ANEOS, Thomas-Fermi-Dirac, etc).  However significant process has been made with first-principles computer simulation \citep{Umemoto06}.  How much iron can post-perovskite accommodate? If this high-pressure silicate phase accommodates a large amount of iron (as suggested by Mao et al \cp{Mao04}), it could affect the size of the core and to a smaller extent the total radius of the planet. (4) What are the thermodynamic properties of all mantle materials, especially the Gruneisen parameter? Post-perovskite has a more sensitive Gruneisen parameter to volume than perovskite, such that the temperature profile of a mantle made mostly of post-perovskite would be cooler than that of perovskite.

\begin{figure}
\includegraphics[scale=0.70]{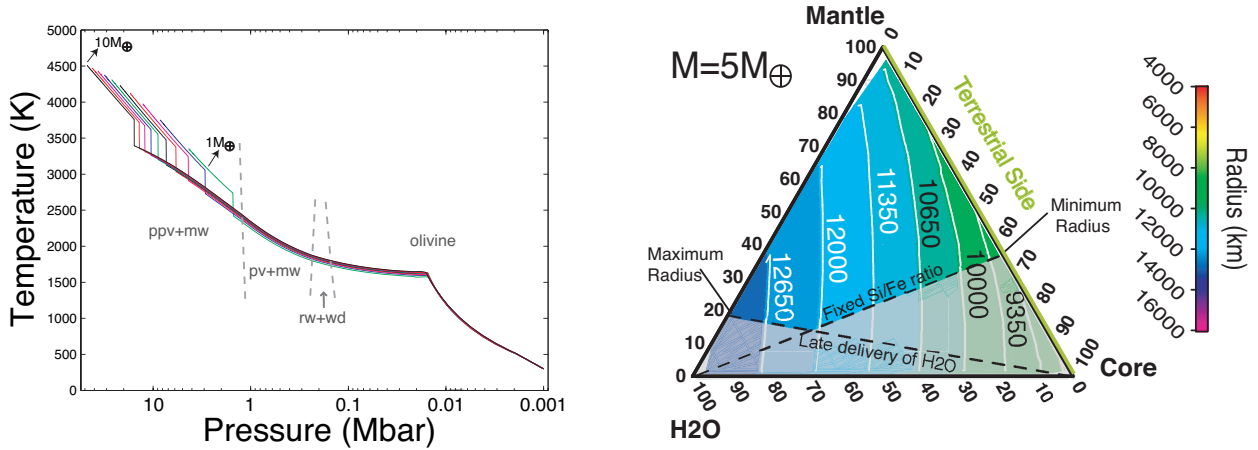}
\caption{(Color online) \emph{Left}: \emph{P-T} profiles of terrestrial super-Earths. The family of planets with 1-10 \me\ have a similar Fe/Si ratio as Earth. The highest internal $P$ is 1.56 Mbar (156 GPa). The different phase transitions in the mantle are shown in dashed lines ranging from olivine (ol), wadsleyite (wd) and ringwoodite (rw), perovskite (pv) and magnesiowusite (wu), and post-perovskite (ppv) and wu. The discontinuities are caused by the boundary layers at the top and bottom of the mantle. The mantles of super-Earths with masses larger than $\sim$4 \me\ are mostly composed of ppv+wu, compared to the dominance of pv+wu on Earth.  \emph{Right}: Ternary Diagram for a 5 \me\ planet. The radius of a planet with each mixture is shown in color with the color bar spanning the radius of the smallest (a 1 \me\ pure Fe planet with $R$ = 5400 km) and largest (a 10 \me\ -- pure H$_2$O planet with $R$ = 16000 km) super-Earth. The shaded region shows the unlikely compositions that can form a super-Earth from solar nebula condensation and secondary accretion constraints. A ternary diagram exists for every planetary mass value.
\label{super}}
\end{figure}

In addition, to infer planetary composition from the $M$ and $R$ data that will be available in the next few years, we need accurate EOSs. Even without errors in the data and structure models, a large number of compositions can fit the same average density (see Figure \ref{super}\emph{b}). The uncertainty in radius from EOS is $\sim$2-3\%, which will be comparable to the precision that powerful space telescopes will yield in follow-up observations (i.e. the \emph{James Webb Space Telescope, JWST}). By reducing the uncertainty in the EOS we make the structure models more accurate and useful. Thus, there is a need for accurate equations of state of solid planetary materials to pressures up to $\sim$ 60 Mbar, the central pressure of the densest and largest (10-\me ) super-Earth.

\section{IV. Conclusions} 
New experiments are now probing states of dense matter that were previously beyond our grasp.  In particular x-ray techniques are allowing us a view into materials that previously had been hidden.  At the same time first principles techniques are allowing accurate determination of EOSs for planetary interest.  Off-Hugoniot experiments of H and He will test these EOS and lead to more accurate models for Jupiter and Saturn.  In addition, new planets are being discovered at an accelerating rate, which will continue to expand the limits of \emph{P-T} space that are of ``planetary'' interest.  Our current era is one of dramatic improving knowledge of, and exciting applications of, the physics of materials at high density.
\\\\
The authors would like to thank their numerous collaborators on these projects.  Please see the references in the text for a fuller account of the science described in this paper.  This work performed under the auspices of the U.S. Department of Energy by Lawrence Livermore National Laboratory under Contract DE-AC52-07NA27344 and at the Los Alamos National Laboratory under Contract DE-AC52-06NA25396. This work is also supported by LDRDs 08-ERI-002, 08-LW-004, and the National Laboratory User Facility program. JJF and BM acknowledge support from NASA and NSF.

%\bibliographystyle{apsrev}
%\bibliography{references}

\end{document}